V.I. Zhaba

Uzhhorod National University, 88000, Uzhhorod, Voloshin Str., 54

# ASYMPTOTICS OF PHASE AND WAVE FUNCTIONS

For single and twochannel nucleon-nucleon scattering the asymptotic form of the phase function for $r_0 \to 0$ were taken into account for the asymptotic behavior of the wave function. Asymptotics of the wave function will not $\sim r_0^{l+1}$, and will have a more complex view and be also determined by the behavior of the potential near the origin. Have examined the cases for nonsingular (weakly singular) and strongly singular potentials. Were the numerical calculations of phase, amplitude and wave functions for the nucleon-nucleon potential Argonne v18. Considered $^1S_0$-, $^3P_0$-, $^3P_1$- states of the *np*- system.

**Key words:** phase shifts, scattering, channel, asymptotic.

В.И. Жаба

Ужгородский национальный университет, 88000, Ужгород, ул. Волошина, 54

# АСИМПТОТИКИ ФАЗОВОЙ И ВОЛНОВОЙ ФУНКЦИЙ

.

Для одно- и двоканального нуклон-нуклонного рассеяния асимптотику фазовой функции при $r_0 \to 0$ учтено для асимптотического поведения волновой функции. Асимптотика волновой функции не будет $\sim r_0^{l+1}$, а будет иметь более сложный вид и определяться также поведением потенциала вблизи начала координат. Рассмотрены случаи для несингулярного (слабо сингулярного) и сильно сингулярного потенциалов. Проведены численные расчеты фазовой, амплитудной и волновой функций для нуклон-нуклонного потенциала Argonne v18. Рассматривались $^1S_0$-, $^3P_0$-, $^3P_1$- состояния для *np*- системы.

**Ключевые слова:** фазовый сдвиг, рассеяние, канал, асимтотика.

В.І. Жаба

Ужгородський національний університет, 88000, Ужгород, вул. Волошина, 54
e-mail: viktorzh@meta.ua

# АСИМПТОТИКИ ФАЗОВОЇ ТА ХВИЛЬОВОЇ ФУНКЦІЙ

Для одно- і двоканального нуклон-нуклонного розсіяння асимптотику фазової функції при $r_0 \to 0$ враховано для асимптотичної поведінки хвильової функції. Асимптотика хвильової функції не буде $\sim r_0^{l+1}$, а матиме складніший вид та визначатиметься також і поведінкою потенціалу поблизу початку координат. Розглянуто випадки для несингулярного (слабо сингулярного) і сильно сингулярного потенціалів. Проведено чисельні розрахунки фазової, амплітудної та хвильової функцій для нуклон-нуклонного потенціалу Argonne v18. Розглядались $^1S_0$ -, $^3P_0$ -, $^3P_1$- стани для *np*- системи.

**Ключові слова:** фазовий зсув, розсіяння, канал, асимтотика,

**Вступ**

Із експериментально спостережуваних величин перерізу розсіяння та енергій переходів отримують у першу чергу інформацію про фази та амплітуди розсіяння, а не про хвильові функції, що є основним об'єктом дослідження при стандартному підході. На експерименті спостерігаються не самі хвильові функції, а певні їх зміни, викликані у результаті взаємодії [1]. Тому необхідно отримати такі співвідношення, що безпосередньо пов'язують фази й амплітуди розсіяння з потенціалом, не знаходячи при цьому хвильові функції.

Точний розв'язок задачі розсіяння із метою обчислення фаз можливе тільки для окремих феноменологічних потенціалів. При виборі реалістичних потенціалів фази розсіяння обчислюються наближено. Це пов'язано з використанням фізичних апроксимацій або з чисельним розрахунком. Вплив вибору чисельного алгоритму на розв'язок задачі розсіяння розглянуто у роботі [2]. Одним із методів знаходження фазових зсувів у задачах для одноканального нуклон-нуклонного розсіяння чи для змішаних станів системи двох нуклонів є метод фазових функцій (МФФ). До методів розв'язування рівняння Шредінгера з метою отримання фаз розсіяння належать: метод послідовних наближень, борнівське наближення, Brysk's- апроксимація.

Дана стаття присвячена розгляду асимптотик фазової та хвильової функцій в підході МФФ.

**Метод фазових функцій: одноканальне розсіяння**

Рівняння Шредінгера для радіальної хвильової функції для задачі розсіяння безспінової частинки з енергією $E$ і орбітальним моментом $l$ на сферично-симетричному потенціалі $V(r)$ має вигляд [1]:

$$u''_l(r) + \left(k^2 - \frac{l(l+1)}{r^2} - U(r)\right)u_l(r) = 0, \quad (1)$$

де $U(r) = 2mV(r)/\hbar^2$ - перенормований потенціал взаємодії, $m$ - приведена маса, $k^2 = 2mE/\hbar^2$ - хвильове число.

Математично метод фазових функцій - це особливий спосіб розв'язку радіального рівняння Шредінгера (1), яке є лінійним диференціальним рівнянням 2-го порядку. Він досить зручний для отримання фаз розсіяння, оскільки по цьому методу не потрібно спочатку обчислювати в широкій області радіальні хвильові функції і потім по їх асимптотикам знаходити ці фази.

Стандартний спосіб обчислення фаз розсіяння - це розв'язок рівняння Шредінгера (1) з асимптотичною граничною умовою. МФФ - це перехід від рівняння Шредінгера до рівняння для фазової функції. Для цього роблять заміну [1, 3]:

$$u_l(r) = A_l(r)[\cos\delta_l(r) \cdot j_l(kr) - \sin\delta_l(r) \cdot n_l(kr)]. \quad (2)$$

Введені дві нові функції $\delta_l(r)$ і $A_l(r)$ мають зміст відповідних фаз розсіяння і констант нормування (амплітуд) хвильових функцій для розсіяння на визначеній послідовності обрізаних потенціалів. $\delta_l(r)$ і $A_l(r)$ називаються відповідно їх фізичному змісту фазовою й амплітудною функцією. Термін "фазова функція" вперше був використаний у роботі Морза і Алліса [4]. Рівняннями для фазової й амплітудної функцій з початковими умовами є:

$$\delta'_l = -\frac{1}{k}U[\cos\delta_l \cdot j_l - \sin\delta_l \cdot n_l]^2, \quad \delta_l(0) = 0; \quad (3)$$

$$A'_l = -\frac{1}{k}A_l U[\cos\delta_l \cdot j_l - \sin\delta_l \cdot n_l] \times \quad (4)$$
$$\times [\sin\delta_l \cdot j_l + \cos\delta_l \cdot n_l], \quad A_l(0) = 1.$$

Фазове рівняння (3) було вперше отримано Друкарєвим [5], а потім незалежно у роботах Бергмана [6] і Колоджеро [7]. Частинний випадок рівняння (6) при $l=0$ був використаний Морзе і Аллісом при дослідженні задачі $S$- розсіяння повільних електронів на атомах [4].

**Асимптотика фазової функції**

Дослідимо поведінку рішень фазового рівняння (3) у залежності від властивостей потенціалу поблизу початку координат. Із-за сингулярності поведінки функції $n_l(kr)$ при $l > 0$ інтегрування рівняння (3) в чисельних розрахунках слід починати від точки $r = r_0 > 0$. Розглянемо асимптотику фазової функції $\delta_l(r_0)$ поблизу початку координат. Для цієї мети зручно переписати фазове рівняння (3) в інтегральній формі:

$$\delta_l(r) = -\frac{1}{k}\int_0^r U(r')[\cos\delta_l \cdot j_l - \sin\delta_l \cdot n_l]^2 dr'. \quad (5)$$

У залежності від поведінки потенціалу на малих відстанях можна розглянути такі два випадки.

1. Потенціал не сингулярний або слабо сингулярний, тобто $r^2 V(r) \to 0$. Враховуючи при малих значеннях аргументів, що величини $\cos\delta_l(r') \approx 1$ і $\sin\delta_l(r') \approx \delta_l(r')$ та сферичні функції Бесселя цілого порядку у виді розкладу тільки з двома першими доданками:

$$j_l(kr) \approx \frac{(kr)^{l+1}}{(2l+1)!!} - \frac{(kr)^{l+3}}{2(2l+3)!!} + ...,$$
$$n_l(kr) \approx -\frac{(2l-1)!!}{(kr)^l} - \frac{(2l-3)!!}{2(kr)^{l-3}} + ...,$$

співвідношення (5) записується у виді

$$\delta_l(r_0) = -\frac{1}{k}\int_0^{r_0} U[\cos\delta_l \cdot j_l - \sin\delta_l \cdot n_l]^2 dr =$$
$$-\frac{1}{k}\int_0^{r_0} U\left[\frac{(kr)^{l+1}}{(2l+1)!!} - \frac{(kr)^{l+3}}{2(2l+3)!!}\right]^2 dr +$$
$$+\frac{2}{k}\int_0^{r_0} U\delta_l\left[\frac{(kr)^{l+1}}{(2l+1)!!} - \frac{(kr)^{l+3}}{2(2l+3)!!}\right] \times \quad (6)$$
$$\times\left[-\frac{(2l-1)!!}{(kr)^l} - \frac{(2l-3)!!}{2(kr)^{l-3}}\right] dr$$
$$-\frac{1}{k}\int_0^{r_0} U\delta_l^2\left[-\frac{(2l-1)!!}{(kr)^l} - \frac{(2l-3)!!}{2(kr)^{l-3}}\right]^2 dr.$$

При достатньо малих $r_0$ другий і третій інтеграли малі в порівнянні з першим. Умови малості їх підінтегрального виразу можна записати при малих $r$ у виді

$$|\delta_l^{(2)}(r_0)| \ll \frac{(kr_0)^{2l+1}}{(2l+1)!!(2l-1)!!},$$
$$|\delta_l^{(3)}(r_0)| \ll \left[\frac{(kr_0)^{2l+1}}{(2l+1)!!(2l-1)!!}\right]^2.$$

Тому при $r_0 \to 0$ в (6) слід обмежитися тільки першим членом у вигляді

$$\delta_l(r_0) \approx -\frac{1}{k}\int_0^{r_0} U\left[\frac{(kr)^{l+1}}{(2l+1)!!} - \frac{(kr)^{l+3}}{2(2l+3)!!}\right]^2 dr.$$

Розкриємо дужки в інтегралі:

$$\delta_l(r_0) \approx -\frac{1}{k}\int_0^{r_0} U\left[\frac{(kr)^{2l+2}}{[(2l+1)!!]^2} - \right.$$
$$\left. -\frac{(kr)^{2l+4}}{(2l+1)!!(2l+3)!!} + \frac{(kr)^{2l+6}}{4[(2l+3)!!]^2}\right] dr,$$

де другий і третій доданки в дужках значно менші від першого, оскільки містять додаткові множники $r^2$ і $r^4$ відповідно, тобто при малих значення ними можна знехтувати:

$$\delta_l(r_0) \approx -\frac{k^{2l+1}}{[(2l+1)!!]^2}\int_0^{r_0} U r^{2l+2} dr. \quad (7)$$

Аналогічне співвідношення отримано у роботі [1]. Крім цього, при несингулярному або слабо сингулярному потенціалі нормування амплітудної функції довільне. Вона може бути нормована на одиницю в будь-якій точці $r_0$, включаючи $r_0=0$.

2. Для сильно сингулярного потенціалу відштовхування $r^2 V(r) \to +\infty$ асимптотика фазової функції при $r_0 \to 0$ буде згідно [1]

$$\delta_l(r_0) \approx -\frac{(kr_0)^{2l+1}}{(2l+1)!!(2l-1)!!}\left[1 - \frac{2l+1}{r_0 U^{1/2}} + O(r_0^2)\right], \quad (8)$$

де враховано перша частина імітує добре відомий ефект твердої відштовхувальної серцевини малого радіусу, а друга - це

поправочний член, який отримується за допомогою наступної підстановки

$$\delta_l(r) = arctg\frac{j_l(kr)}{n_l(kr)} + \Delta_l(r), \ r \to 0. \quad (9)$$

### Асимптотика хвильової функції

Знайдемо асимптотику для хвильової функції (2) для одноканального розсіяння. З урахуванням асимптотик сферичних функцій Бесселя, матимемо

$$u_l(r_0) = A_l(r_0)\left[\cos\delta_l(r_0)\cdot j_l - \sin\delta_l(r_0)\cdot n_l\right] \approx$$
$$\approx \frac{(kr_0)^{l+1}}{(2l+1)!!} - \frac{(kr_0)^{l+3}}{2(2l+3)!!} + \delta_l(r_0)\left[\frac{(2l-1)!!}{(kr_0)^l} + \frac{(2l-3)!!}{2(kr_0)^{l-3}}\right].$$

Оскільки нам відомі асимптотики (7) і (8) для фазової функції, то їх можна використати для запису асимптотики хвильової функції поблизу початку координат. Якщо врахувати тільки перші члени асимптотичного розкладу сферичних функцій Бесселя, то при $r_0 \to 0$ асимптотика для хвильової функції буде:

1) для несингулярного або слабо сингулярного потенціалу

$$u_l(r_0) \approx \frac{(kr_0)^{l+1}}{(2l+1)!!} - \frac{(2l-1)!!}{[(2l+1)!!]^2}k^{l+1}r_0^{-l}\int_0^{r_0}Ur^{2l+2}dr. \quad (10)$$

2) для сильно сингулярного потенціалу відштовхування

$$u_l(r_0) \approx \frac{k^{l+1}}{(2l-1)!!}\frac{r_0^l}{U^{1/2}(r_0)}. \quad (11)$$

Як видно з формул (10) і (11), асимптотика хвильової функції не буде $u(r_0) \sim r_0^{l+1}$, а матиме складніший вид та визначатиметься також і поведінкою потенціалу поблизу початку координат.

### Метод фазових функцій: двоканальне розсіяння

Для знаходження фазових зсувів змішаних станів системи двох нуклонів потрібно розв'язувати зв'язану систему рівнянь Шредінгера з тензорним змішуванням [1,8]:

$$\begin{cases} u'' + \left[k^2 - \frac{J(J-1)}{r^2} - U_1\right]u = U_3 w, \\ w'' + \left[k^2 - \frac{(J+2)(J+1)}{r^2} - U_2\right]w = U_3 u, \end{cases} \quad (12)$$

Після підстановки хвильових функцій

$$\begin{cases} u = A\left[\cos\delta_1 \cdot j_1 - \sin\delta_1 \cdot n_1\right], \\ w = B\left[\cos\delta_2 \cdot j_2 - \sin\delta_2 \cdot n_2\right] \end{cases} \quad (13)$$

в (12) отримуємо систему чотирьох нелінійних зв'язаних диференціальних рівнянь 1-го порядку для функцій фазових й амплітудних у такому виді

$$\begin{cases} \delta_1' = -\frac{U_1}{k_1}P_1^2 - \frac{U_3 tg\varepsilon}{k_1}P_1 P_2; \\ \delta_2' = -\frac{U_2}{k_2}P_2^2 - \frac{U_3}{k_2 tg\varepsilon}P_2 P_1; \\ A' = -\frac{U_1 A}{k_1}P_1 Q_1 - \frac{U_3 A tg\varepsilon}{k_1}P_2 Q_1; \\ B' = -\frac{U_2 A tg\varepsilon}{k_2}P_2 Q_2 - \frac{U_3 A}{k_2}P_1 Q_2; \end{cases} \quad (14)$$

де для спрощення запису введені наступні позначення

$$\begin{aligned} P_1 &= \cos\delta_1 \cdot j_1 - \sin\delta_1 \cdot n_1; \\ P_2 &= \cos\delta_2 \cdot j_2 - \sin\delta_2 \cdot n_2; \\ Q_1 &= \sin\delta_1 \cdot j_1 + \cos\delta_1 \cdot n_1; \\ Q_2 &= \sin\delta_2 \cdot j_2 + \cos\delta_2 \cdot n_2. \end{aligned}$$

В формулах системи (14) параметр змішування амплітуд розсіяння і вронскіани розв'язків вільного рівняння Шредінгера рівні

$$\begin{cases} tg\varepsilon = \frac{B}{A}; \\ k_1 = j_1 \cdot n_1' - n_1 \cdot j_1'; \\ k_2 = j_2 \cdot n_2' - n_2 \cdot j_2'. \end{cases}$$

Враховуючи асимптотику сферичних функцій Бесселя, отримаємо асимптотики для фаз розсіяння $\delta_1$ і $\delta_2$ для слабо сингулярного потенціалу:

$$\begin{cases} \delta_1^/(r_0) \approx -\dfrac{k_1^{2l_1+1}}{[(2l_1+1)!!]^2}\int_0^{r_0} U_1 r^{2l_1+2} dr - \\ \qquad -\dfrac{k_1^{l_1} k_2^{l_2+1}}{(2l_1+1)!!(2l_2+1)!!}\int_0^{r_0} U_3 tg\varepsilon\, r^{l_1+l_2+2} dr; \\ \delta_2^/(r_0) \approx -\dfrac{k_2^{2l_2+1}}{[(2l_2+1)!!]^2}\int_0^{r_0} U_2 r^{2l_2+2} dr - \\ \qquad -\dfrac{k_1^{l_1+1} k_2^{l_2}}{(2l_1+1)!!(2l_2+1)!!}\int_0^{r_0} \dfrac{U_3}{tg\varepsilon} r^{l_1+l_2+2} dr; \end{cases} \quad (15)$$

і для сильно сингулярного потенціалу:

$$\begin{cases} \delta_1^/(r_0) \approx -\dfrac{(k_1 r_0)^{2l_1+1}}{(2l_1+1)!!(2l_1-1)!!}\left[1-\dfrac{2l_1+1}{r_0 U_1^{1/2}}\right] - \\ \qquad -V_3 tg\varepsilon \dfrac{k_1^{3l_1+1} k_2^{l_2} r^{3l_1+l_2+2}}{[(2l_1+1)!!]^2(2l_1-1)!!(2l_2-1)!!}[1+O_1]; \\ \delta_2^/(r_0) \approx -\dfrac{(k_2 r_0)^{2l_2+1}}{(2l_2+1)!!(2l_2-1)!!}\left[1-\dfrac{2l_2+1}{r_0 U_2^{1/2}}\right] - \\ \qquad -\dfrac{V_3}{tg\varepsilon} \dfrac{k_1^{l_1} k_2^{3l_2+1} r^{l_1+3l_2+2}}{[(2l_2+1)!!]^2(2l_1-1)!!(2l_2-1)!!}[1+O_2]; \end{cases} \quad (16)$$

де поправочні доданки $O_1(r_0^2)$ і $O_2(r_0^2)$ визначають за допомогою виразів виду (9). Крім асимтотик для фазових функцій, необхідно знаходити і асимптотику для параметру змішування амплітуд розсіяння.

Систему рівнянь (12) можна розв'язати за допомогою параметризацій [1] Мак-Хейла-Телера, Блатта-Біденхарна, Стаппа чи Матвєєнка-Пономарьова-Файфмана. Асимптотики фаз для цих параметризацій набуватимуть значно складнішого виду, ніж (15) або (16).

### Чисельні розрахунки

Вказані співвідношення для асимтотик фазової та хвильової функції застосуємо для конкретних потенціалів у задачах одноканального розсіяння.

Для несингулярного потенціалу $U = U_0 e^{-a/r}$ знайдено асимптотику хвильової функції (10), причому

$$u_l(r_0) \approx \dfrac{(kr_0)^{l+1}}{(2l+1)!!} - \dfrac{(2l-1)!!}{[(2l+1)!!]^2} k^{l+1} r_0^{-l} \int_0^{r_0} U r^{2l+2} dr = \\ = u_l^{(1)}(r_0) + u_l^{(2)}(r_0), \quad r_0 \to 0. \quad (17)$$

Для $l=0$; $k=10$; $U_0=7,5$; $a=0,2$ отримані значення компонент асимптотики хвильової функції приведено в Таблиці 1. Як бачимо, що друга частина асимптотики $u_l^{(2)}(r_0)$, яка залежить від форми потенціалу взаємодії, дає певний вклад в повну асимптотику хвильової функції $u_l(r_0)$.

Таблиця 1
**Асимптотика хвильової функції для несингулярного потенціалу**

| $r_0$ | $u_l^{(1)}(r_0)$ | $u_l^{(2)}(r_0)$ | $u_l(r_0)$ |
|---|---|---|---|
| 0,001 | 0,0100 | -5,09E-97 | 0,0100 |
| 0,008 | 0,0800 | -1,85E-17 | 0,0800 |
| 0,015 | 0,1500 | -2,39E-11 | 0,1500 |
| 0,022 | 0,2200 | -7,02E-09 | 0,2200 |
| 0,029 | 0,2900 | -1,75E-07 | 0,2900 |
| 0,036 | 0,3600 | -1,47E-06 | 0,3600 |
| 0,043 | 0,4300 | -6,91E-06 | 0,4300 |
| 0,05 | 0,5000 | -2,27E-05 | 0,5000 |
| 0,057 | 0,5700 | -5,91E-05 | 0,5699 |
| 0,064 | 0,6400 | -1,30E-04 | 0,6399 |
| 0,071 | 0,7100 | -2,55E-04 | 0,7097 |

Розглянемо $^1S_0$-, $^3P_0$-, $^3P_1$- стани для $np$- системи, де маси нуклонів було вибрано такими: $M_p$=938,27231 MeV; $M_n$=939,56563 MeV. На Рис. 1-3 приведено розраховані по МФФ фазова й амплітудна функції для потенціалу нуклон-нуклонної взаємодії Argonne v18 [9]. Фазові зсуви вказано у градусах. На рисунках також зображена і хвильова функція, розрахована по формулі (2). Враховано асимптотики в залежності від потенціалу нуклон-нуклонної взаємодії. Фазова функція виходить на асимптотику, і фаза розсіяння для даних спінових станів співпадає з розрахунками оригінальної роботи [9].

Доцільним є розрахунок фазових зсувів для двоканального розсіяння, враховуючи отримані асимптотики виду (15) і (16). Наприклад, цілком можливо застосувати при розрахунку фазових зсувів і параметру змішування зв'язаних станів $^3S_1$-$^3D_1$ або $^3P_2$-$^3F_2$.

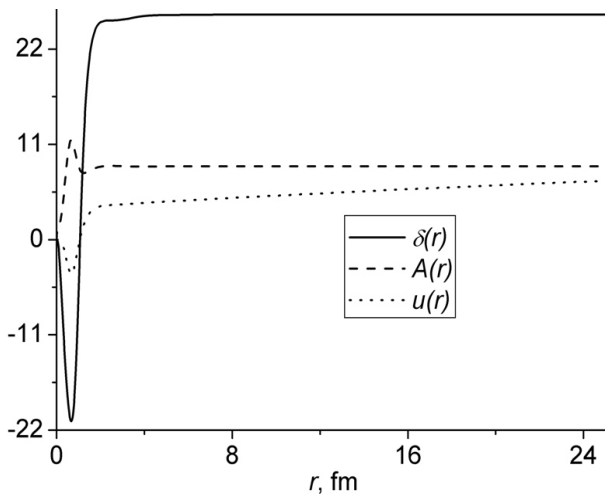

Рис. 1. Фазова, амплітудна і хвильова функції для $^1S_0$ – стану

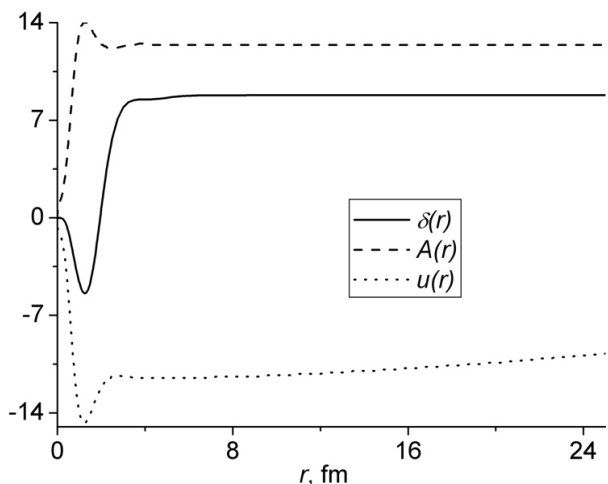

Рис. 2. Фазова, амплітудна і хвильова функції для $^3P_0$ - стану

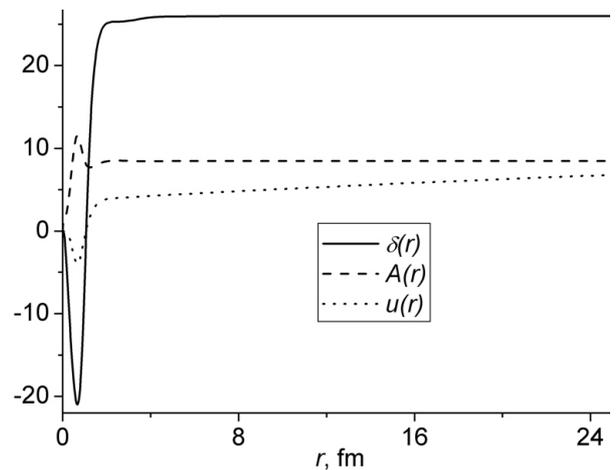

Рис. 3. Фазова, амплітудна і хвильова функції для $^3P_1$- стану

**Висновки**

Асимптотика фазової функції $\delta_l(r_0)$ для одно- і двоканального нуклон-нуклонного розсіяння визначається формою потенціалу.

Асимптотику фазової функції при $r_0 \to 0$ враховано для асимптотичної поведінки хвильової функції.

Асимптотики фазової та хвильової функцій розглянуто для несингулярного (слабо сингулярного) і сильно сингулярного потенціалів.

У рамках МФФ проведено чисельні розрахунки фазової, амплітудної та хвильової функцій для потенціалу нуклон-нуклонної взаємодії Argonne v18. Безпосередньо розглядались $^1S_0$ -, $^3P_0$ -, $^3P_1$- стани для $np$- системи.

СПИСОК ВИКОРИСТАНОЇ ЛІТЕРАТУРИ